\documentclass[11pt]{article}
\usepackage{xypic}
\usepackage{mathrsfs}
\usepackage{amssymb}
\usepackage{amsfonts}
\usepackage{amsmath}

\def\ben{\begin{enumerate}}
\def\een{\end{enumerate}}

\def\bit{\begin{itemize}}
\def\eit{\end{itemize}}

\def\0{\leqno}

\input xy
\xyoption{all}
\input{tcilatex}
\begin{document}

\begin{center}
\textbf{INTERIOR AND EXTERIOR DIFFERENTIAL SYSTEMS }

\textbf{FOR LIE ALGEBROIDS}

\bigskip \ \ \ \ 


\textbf{{CONSTANTIN M. ARCU\c{S} }}

\begin{tabular}{c}
Secondary School \textquotedblleft CORNELIUS RADU\textquotedblright , \\
R\u{a}dine\c{s}ti Village, 217196, Gorj County, Rom\^{a}nia \\
e-mail: c\_arcus@yahoo.com, c\_arcus@radinesti.ro%
\end{tabular}
\end{center}

\ \ \ \ \bigskip


\ \ \textit{In memory of my uncle }

\ \ \textit{Prof. Dr. Gheorghe RADU}

\ \ \ \

\begin{abstract}
A theorem of Maurer-Cartan type for Lie algebroids is presented. Suppose
that any vector subbundle of a Lie algebroid is called \emph{interior
differential system (IDS)} for that Lie algebroid. A theorem of Cartan type
is obtained. Extending the classical notion of \emph{exterior differential
system (EDS)} to Lie algebroids, a theorem of Cartan type is obtained. \ \
\end{abstract}





\ \

\ \ \ 

\section{Introduction}

Using the exterior differential calculus for Lie algebroids (see: $\left[ 2,6%
\right] $) the structure equations of Maurer-Cartan type are established.

Using the \emph{Cartan's moving frame method}, there exists the following

\textbf{Theorem }(E. Cartan) \emph{If }$N\in \left\vert \mathbf{Man}%
_{n}\right\vert $\emph{\ is a Riemannian manifold and }$X_{\alpha
}=X_{\alpha }^{i}\frac{\partial }{\partial x^{i}},$\emph{\ }$\alpha \in
\overline{1,n}$\emph{\ is an ortonormal moving frame, then there exists a
colection of }$1$\emph{-forms }$\Omega _{\beta }^{\alpha },~\alpha ,\beta
\in \overline{1,n}$\emph{\ uniquely defined by the requirements}%
\begin{equation*}
\Omega _{\beta }^{\alpha }=-\Omega _{\alpha }^{\beta }
\end{equation*}%
\emph{and }%
\begin{equation*}
d^{F}\Theta ^{\alpha }=\Omega _{\beta }^{\alpha }\wedge \Theta ^{\beta
},~\alpha \in \overline{1,n}
\end{equation*}%
\emph{where }$\left\{ \Theta ^{\alpha },\alpha \in \overline{1,n}\right\} $%
\emph{\ is the coframe.} (see $\left[ 5\right] ,$ p. 151)

We know that an $r$\emph{-dimensional distribution on a manifold }$N$ is a
mapping $\mathcal{D}$ defined on $N,$ which assignes to each point $x$ of $N$
an $r$-dimensional linear subspace $\mathcal{D}_{x}$ of $T_{x}N.$ A vector
fields $X$ belongs to $\mathcal{D}$ if we have $X_{x}\in \mathcal{D}_{x}$
for each $x\in N.$ When this happens we write $X\in \Gamma \left( \mathcal{D}%
\right) .$

The distribution $\mathcal{D}$ on a manifold $N$ is said to be \emph{%
differentiable} if for any $x\in N$ there exists $r$ differentiable linearly
independent vector fields $X_{1},...,X_{r}\in \Gamma \left( \mathcal{D}%
\right) $ in a neighborhood of $x.$ The distribution $\mathcal{D}$ is said
to be \emph{involutive }if for all vector fields $X,Y\in \Gamma \left(
\mathcal{D}\right) $ we have $\left[ X,Y\right] \in \Gamma \left( \mathcal{D}%
\right) .$

Extending the notion of distribution we obtain the definition of an \emph{IDS%
} of a Lie algebroid. A characterization of the ivolutivity of an \emph{IDS}
in a result of Cartan type is presented in \emph{Theorem 4.7.}

In the classical theory we have the following

\textbf{Theorem }(Frobenius)\textbf{\ }\emph{The distribution }$\mathcal{D}$%
\emph{\ is involutive if and only if for each }$x\in N$\emph{\ there exists
a neighborhood }$U$\emph{\ and }$n-r$\emph{\ linearly independent }$1$\emph{%
-forms }$\Theta ^{r+1},...,\Theta ^{n}$\emph{\ on }$U$\emph{\ which vanish
on }$\mathcal{D}$\emph{\ and satisfy the condition }%
\begin{equation*}
d^{F}\Theta ^{\alpha }=\Sigma _{\beta \in \overline{r+1,p}}\Omega _{\beta
}^{\alpha }\wedge \Theta ^{\beta },~\alpha \in \overline{r+1,n}.
\end{equation*}%
\emph{for suitable }$1$\emph{-forms }$\Omega _{\beta }^{\alpha },~\alpha
,\beta \in \overline{r+1,n}.$(see $\left[ 4\right] ,$ p. 58)

This paper studies the intersection between the geometry of Lie algebroids
and some aspects of \emph{EDS}. In the classical sense, an \emph{EDS} is a
pair $(M,\mathcal{E})$ consisting of a smooth manifold $M$ and a
homogeneous, differentially closed ideal $\mathcal{E}\subseteq \Omega ^{\ast
}\left( M\right) $ in the algebra of smooth differential forms on $M$. ( see
$[1,3])$ Using the notion of \emph{EDS} of an arbitrary Lie algebroid $%
\left( \left( F,\nu ,N\right) ,\left[ ,\right] _{F},\left( \rho
,Id_{N}\right) \right) $ we obtained a new result of Cartan type in the
\emph{Theorem 5.1. }

In the particular case of standard Lie algebroid $\left[ \left( TM,\tau
_{M},M\right) ,\left[ ,\right] _{TM},\left( Id_{TM},Id_{M}\right) \right] $
there are obtained similar results those for distributions.

We know that a submanifold $S$ of $N$ is said to be \emph{integral manifold }%
for the dstribution\emph{\ }$\mathcal{D}$ if for every point $x\in N,$ $%
\mathcal{D}_{x}$ coincides with $T_{x}S.$ The distribution $\mathcal{D}$ is
said to be\emph{\ integrable} if for each point $x\in N$ there exists an
integral manifold of $\mathcal{D}$ containing $x.$

As a distribution $\mathcal{D}$ is involutive if and only if it is
integrable, then the study of the integral manifolds of an \emph{IDS} or
\emph{EDS} is a new direction by research.









\section{Preliminaries}

In general, if $\mathcal{C}$ is a category, then we denote $\left\vert
\mathcal{C}\right\vert $ the class of objects and for any $A,B{\in }%
\left\vert \mathcal{C}\right\vert $, we denote $\mathcal{C}\left( A,B\right)
$ the set of morphisms of $A$ source and $B$ target. Let$\mathbf{~Liealg},~%
\mathbf{Mod,~}$and $\mathbf{B}^{\mathbf{v}}$ be the category of Lie
algebras, modules and vector bundles respectively.

We know that if $\left( E,\pi ,M\right) \in \left\vert \mathbf{B}^{\mathbf{v}%
}\right\vert ,$ $\Gamma \left( E,\pi ,M\right) =\left\{ u\in \mathbf{Man}%
\left( M,E\right) :u\circ \pi =Id_{M}\right\} $ and $\mathcal{F}\left(
M\right) =\mathbf{Man}\left( M,\mathbb{R}\right) ,$ then $\left( \Gamma
\left( E,\pi ,M\right) ,+,\cdot \right) $ is a $\mathcal{F}\left( M\right) $%
-module. Aditionally, if $\left( E,\pi ,M\right) \in \left\vert \mathbf{B}^{%
\mathbf{v}}\right\vert $ so that $M$ is paracompact and if $A\subseteq M$ is
closed, then for any section $u$ over $A$ it exists $\tilde{u}\in $ $\Gamma
\left( E,\pi ,M\right) $ so that $\tilde{u}_{|A}=u.$ In the following, we
consider only vector bundles with paracompact base.

We know that a Lie algebroid is a vector bundle $\left( F,\nu ,N\right) \in
\left\vert \mathbf{B}^{\mathbf{v}}\right\vert $ so that there exists
\begin{equation*}
\begin{array}{c}
\left( \rho ,Id_{N}\right) \in \mathbf{B}^{\mathbf{v}}\left( \left( F,\nu
,N\right) ,\left( TN,\tau _{N},N\right) \right)%
\end{array}%
\end{equation*}%
and an operation
\begin{equation*}
\begin{array}{ccc}
\Gamma \left( F,\nu ,N\right) \times \Gamma \left( F,\nu ,N\right) & ^{%
\underrightarrow{\,\left[ ,\right] _{F}\,}} & \Gamma \left( F,\nu ,N\right)
\\
\left( u,v\right) & \longmapsto & \left[ u,v\right] _{F}%
\end{array}%
\end{equation*}%
with the following properties:

\begin{itemize}
\item[$LA_{1}$.] the equality holds good
\begin{equation*}
\begin{array}{c}
\left[ u,f\cdot v\right] _{F}=f\left[ u,v\right] _{F}+\Gamma \left( \rho
,Id_{N}\right) \left( u\right) f\cdot v,%
\end{array}%
\end{equation*}%
for all $u,v\in \Gamma \left( F,\nu ,N\right) $ and $f\in \mathcal{F}\left(
N\right) ,$

\item[$LA_{2}$.] the $4$-tuple $\left( \Gamma \left( F,\nu ,N\right)
,+,\cdot ,\left[ ,\right] _{F}\right) $ is a Lie $\mathcal{F}\left( N\right)
$-algebra$,$

\item[$LA_{3}$.] the $\mathbf{Mod}$-morphism $\Gamma \left( \rho
,Id_{N}\right) $ is a $\mathbf{LieAlg}$-morphism of $\left( \Gamma \left(
F,\nu ,N\right) ,+,\cdot ,\left[ ,\right] _{F}\right) $ source and $\left(
\Gamma \left( TN,\tau _{N},N\right) ,+,\cdot ,\left[ ,\right] _{TN}\right) $
target.
\end{itemize}

Let $\left( \left( F,\nu ,N\right) ,\left[ ,\right] _{F,},\left( \rho
,Id_{N}\right) \right) $ be a Lie algebroid.

\begin{itemize}
\item Locally, for any $\alpha ,\beta \in \overline{1,p},$ we set $\left[
t_{\alpha },t_{\beta }\right] _{F}=L_{\alpha \beta }^{\gamma }t_{\gamma }.$
We easily obtain that $L_{\alpha \beta }^{\gamma }=-L_{\beta \alpha
}^{\gamma },~$for any $\alpha ,\beta ,\gamma \in \overline{1,p}.$
\end{itemize}

The real local functions $\left\{ L_{\alpha \beta }^{\gamma },~\alpha ,\beta
,\gamma \in \overline{1,p}\right\} $ are called the \emph{structure
functions.}

\begin{itemize}
\item We assume that $\left( F,\nu ,N\right) $ is a vector bundle with type
fibre the real vector space $\left( \mathbb{R}^{p},+,\cdot \right) $ and
structure group a Lie subgroup of $\left( \mathbf{GL}\left( p,\mathbb{R}%
\right) ,\cdot \right) .$ We denote $(x^{i},z^{\alpha })$ the canonical
local coordinates on $(F,\nu ,N),$ where $i{\in }\overline{1,n}$, $\alpha
\in \overline{1,p}.$
\end{itemize}

Consider
\begin{equation*}
\left( x^{i},z^{\alpha }\right) \longrightarrow \left( x^{i^{\prime
}},z^{\alpha
{\acute{}}%
}\right)
\end{equation*}%
a change of coordinates on $\left( F,\nu ,N\right) $. Then the coordinates $%
z^{\alpha }$ change to $z^{\alpha
{\acute{}}%
}$ according to the rule:
\begin{equation*}
\begin{array}{c}
z^{\alpha
{\acute{}}%
}=\Lambda _{\alpha }^{\alpha
{\acute{}}%
}z^{\alpha }.%
\end{array}%
\leqno(2.1)
\end{equation*}

\begin{itemize}
\item If $z^{\alpha }t_{\alpha }\in \Gamma \left( F,\nu ,N\right) $ is
arbitrary, then
\begin{equation*}
\begin{array}[t]{l}
\left[ \Gamma \left( \rho ,Id_{N}\right) \left( z^{\alpha }t_{\alpha
}\right) f\right] \left( x\right) =\vspace*{1mm}\left( \rho _{\alpha
}^{i}z^{\alpha }\frac{\partial f}{\partial \varkappa ^{\tilde{\imath}}}%
\right) \left( x\right)%
\end{array}%
\leqno(2.2)
\end{equation*}%
for any $f\in \mathcal{F}\left( N\right) $ and $x\in N.$
\end{itemize}

The coefficients $\rho _{\alpha }^{i}$ change to $\rho _{\alpha
{\acute{}}%
}^{i%
{\acute{}}%
}$ according to the rule:
\begin{equation*}
\begin{array}{c}
\rho _{\alpha
{\acute{}}%
}^{i%
{\acute{}}%
}=\Lambda _{\alpha
{\acute{}}%
}^{\alpha }\rho _{\alpha }^{i}\displaystyle\frac{\partial x^{i%
{\acute{}}%
}}{\partial x^{i}},%
\end{array}%
\leqno(2.3)
\end{equation*}%
where
\begin{equation*}
\left\Vert \Lambda _{\alpha
{\acute{}}%
}^{\alpha }\right\Vert =\left\Vert \Lambda _{\alpha }^{\alpha
{\acute{}}%
}\right\Vert ^{-1}.
\end{equation*}

The following equalities hold good:%
\begin{equation*}
\begin{array}{c}
\displaystyle\left( \rho _{\alpha }^{i}\frac{\partial }{\partial x^{i}}%
\right) \left( f\right) =\rho _{\alpha }^{i}\frac{\partial f}{\partial x^{i}}%
,\forall f\in \mathcal{F}\left( N\right) .%
\end{array}%
\leqno(2.4)
\end{equation*}%
and\emph{\ }%
\begin{equation*}
\begin{array}{c}
\displaystyle L_{\alpha \beta }^{\gamma }\cdot \rho _{\gamma }^{k}=\rho
_{\alpha }^{i}\cdot \frac{\partial \rho _{\beta }^{k}}{\partial x^{i}}-\rho
_{\beta }^{j}\cdot \frac{\partial \rho _{\alpha }^{k}}{\partial x^{j}}.%
\end{array}%
\leqno(2.5)
\end{equation*}

\section{Interior Differential Systems}

Let $\left( \left( F,\nu ,N\right) ,\left[ ,\right] _{F},\left( \rho
,Id_{N}\right) \right) $ be a Lie algebroid.

\textbf{Definition 3.1 }Any vector subbundle $\left( E,\pi ,N\right) $ of
the vector bundle $\left( F,\nu ,N\right) $ will be called \emph{interior
differential system (IDS) of the Lie algebroid }%
\begin{equation*}
\left( \left( F,\nu ,N\right) ,\left[ ,\right] _{F},\left( \rho
,Id_{N}\right) \right) .
\end{equation*}

\textit{Remark 3.1 }If $\left( E,\pi ,M\right) $ is an \emph{IDS} of the Lie
algebroid
\begin{equation*}
\left( \left( F,\nu ,N\right) ,\left[ ,\right] _{F},\left( \rho
,Id_{N}\right) \right) ,
\end{equation*}%
then we obtain a vector subbundle $\left( E^{0},\pi ^{0},N\right) $ of the
dual vector bundle $\left( \overset{\ast }{F},\overset{\ast }{\nu },N\right)
$ so that
\begin{equation*}
\Gamma \left( E^{0},\pi ^{0},N\right) \overset{put}{=}\left\{ \Omega \in
\Gamma \left( \overset{\ast }{F},\overset{\ast }{\nu },N\right) :\Omega
\left( S\right) =0,~\forall S\in \Gamma \left( E,\pi ,N\right) \right\} .
\end{equation*}

The vector subbundle $\left( E^{0},\pi ^{0},N\right) $ will be called \emph{%
the annihilator vector subbundle of the IDS }$\left( E,\pi ,N\right) .$

\textbf{Proposition 3.1 }\emph{If }$\left( E,\pi ,N\right) $\emph{\ is an
IDS of the Lie algebroid }%
\begin{equation*}
\left( \left( F,\nu ,N\right) ,\left[ ,\right] _{F},\left( \rho
,Id_{N}\right) \right)
\end{equation*}%
\emph{so that }$\Gamma \left( E,\pi ,N\right) =\left\langle
S_{1},...,S_{r}\right\rangle $\emph{, then it exists }$\Theta
^{r+1},...,\Theta ^{p}\in \Gamma \left( \overset{\ast }{F},\overset{\ast }{%
\nu },N\right) $\emph{\ linearly independent so that }$\Gamma \left(
E^{0},\pi ^{0},N\right) =\left\langle \Theta ^{r+1},...,\Theta
^{p}\right\rangle .$

\textbf{Definition 3.2 }The \emph{IDS} $\left( E,\pi ,N\right) $ of the Lie
algebroid
\begin{equation*}
\left( \left( F,\nu ,N\right) ,\left[ ,\right] _{F},\left( \rho
,Id_{N}\right) \right)
\end{equation*}%
will be called \emph{involutive} if $\left[ S,T\right] _{F}\in \Gamma \left(
E,\pi ,N\right) ,~$for any $S,T\in \Gamma \left( E,\pi ,N\right) .$

\textbf{Proposition 3.2 }\emph{If }$\left( E,\pi ,N\right) $\emph{\ is an
IDS of the Lie algebroid }%
\begin{equation*}
\left( \left( F,\nu ,N\right) ,\left[ ,\right] _{F},\left( \rho ,\eta
\right) \right)
\end{equation*}%
\emph{and }$\left\{ S_{1},...,S_{r}\right\} $\emph{\ is a base of the }$%
\mathcal{F}\left( M\right) $\emph{-submodule }$\left( \Gamma \left( E,\pi
,N\right) ,+,\cdot \right) $\emph{\ then }$\left( E,\pi ,N\right) $\emph{\
is involutive if and only if }$\left[ S_{a},S_{b}\right] _{F}\in \Gamma
\left( E,\pi ,N\right) ,~$for any $a,b\in \overline{1,r}.$

\section{Exterior differential calculus}

Let $\left( \left( F,\nu ,N\right) ,\left[ ,\right] _{F},\left( \rho
,Id_{N}\right) \right) $ be a Lie algebroid.

We denote $\Lambda ^{q}\left( F,\nu ,N\right) $ the set of \emph{%
differential forms of degree }$q.$ If
\begin{equation*}
\Lambda \left( F,\nu ,N\right) =\underset{q\geq 0}{\oplus }\Lambda
^{q}\left( F,\nu ,N\right) ,
\end{equation*}%
then we obtain \emph{the exterior differential algebra} $\left( \Lambda
\left( F,\nu ,N\right) ,+,\cdot ,\wedge \right) .$

\textbf{Definition 4.1 }For any $z\in \Gamma \left( F,\nu ,N\right) $, the
application
\begin{equation*}
\begin{array}{c}
\begin{array}{rcl}
\Lambda \left( F,\nu ,N\right) & ^{\underrightarrow{~\ \ L_{z}~\ \ }} &
\Lambda \left( F,\nu ,N\right)%
\end{array}%
,%
\end{array}%
\end{equation*}%
defined by%
\begin{equation*}
\begin{array}{c}
L_{z}\left( f\right) =\left[ \Gamma \left( \rho ,Id_{N}\right) z\right]
\left( f\right) ,%
\end{array}%
\end{equation*}%
for any $f\in \mathcal{F}\left( N\right) $ and
\begin{equation*}
\begin{array}{cl}
L_{z}\omega \left( z_{1},...,z_{q}\right) & =\left[ \Gamma \left( \rho
,Id_{N}\right) z\right] \left( \omega \left( \left( z_{1},...,z_{q}\right)
\right) \right) \\
& -\overset{q}{\underset{i=1}{\tsum }}\omega \left( \left( z_{1},...,\left[
z,z_{i}\right] _{F},...,z_{q}\right) \right) ,%
\end{array}%
\end{equation*}%
for any $\omega \in \Lambda ^{q}\mathbf{\ }\left( F,\nu ,N\right) $ and $%
z_{1},...,z_{q}\in \Gamma \left( F,\nu ,N\right) ,$ is called \emph{the
covariant Lie derivative with respect to the section }$z.$

\textbf{Theorem 4.1 }\emph{If }$z\in \Gamma \left( F,\nu ,N\right) ,$ $%
\omega \in \Lambda ^{q}\left( F,\nu ,N\right) $\emph{\ and }$\theta \in
\Lambda ^{r}\left( F,\nu ,N\right) $\emph{, then}%
\begin{equation*}
\begin{array}{c}
L_{z}\left( \omega \wedge \theta \right) =L_{z}\omega \wedge \theta +\omega
\wedge L_{z}\theta .%
\end{array}%
\leqno(4.1)
\end{equation*}

\textbf{Definition 4.2 }For any $z\in \Gamma \left( F,\nu ,N\right) $, the
application%
\begin{equation*}
\begin{array}{rcl}
\Lambda \left( F,\nu ,N\right) & ^{\underrightarrow{\ \ i_{z}\ \ }} &
\Lambda \left( F,\nu ,N\right) \\
\Lambda ^{q}\left( F,\nu ,N\right) \ni \omega & \longmapsto & i_{z}\omega
\in \Lambda ^{q-1}\left( F,\nu ,N\right) ,%
\end{array}%
\end{equation*}%
defined by $i_{z}f=0,$ for any $f\in \mathcal{F}\left( N\right) $ and
\begin{equation*}
\begin{array}{c}
i_{z}\omega \left( z_{2},...,z_{q}\right) =\omega \left(
z,z_{2},...,z_{q}\right) ,%
\end{array}%
\end{equation*}%
for any $z_{2},...,z_{q}\in \Gamma \left( F,\nu ,N\right) $, is called the
\emph{interior product associated to the section}~$z.$\bigskip

\textbf{Theorem 4.2 }\emph{If }$z\in \Gamma \left( F,\nu ,N\right) $\emph{,
then for any }$\omega \in $\emph{\ }$\Lambda ^{q}\left( F,\nu ,N\right) $%
\emph{\ and }$\theta \in $\emph{\ }$\Lambda ^{r}\left( F,\nu ,N\right) $%
\emph{\ we obtain}
\begin{equation*}
\begin{array}{c}
i_{z}\left( \omega \wedge \theta \right) =i_{z}\omega \wedge \theta +\left(
-1\right) ^{q}\omega \wedge i_{z}\theta .%
\end{array}%
\leqno(4.2)
\end{equation*}

\textbf{Theorem 4.3 }\emph{For any }$z,v\in \Gamma \left( F,\nu ,N\right) $%
\emph{\ we obtain}%
\begin{equation*}
\begin{array}{c}
L_{v}\circ i_{z}-i_{z}\circ L_{v}=i_{\left[ z,v\right] _{F}}.%
\end{array}%
\leqno(4.3)
\end{equation*}

\noindent

\textbf{Theorem 4.4 }\emph{The application }%
\begin{equation*}
\begin{array}{c}
\begin{array}{ccc}
\Lambda ^{q}\mathbf{\ }\left( F,\nu ,N\right) & ^{\underrightarrow{\,\
d^{F}\,\ }} & \Lambda ^{q+1}\mathbf{\ }\left( F,\nu ,N\right) \\
\omega & \longmapsto & d\omega%
\end{array}%
\end{array}%
\end{equation*}%
\emph{defined by}%
\begin{equation*}
\begin{array}{c}
d^{F}f\left( z\right) =\Gamma \left( \rho ,Id_{N}\right) \left( z\right) f,%
\end{array}%
\end{equation*}%
\emph{for any }$z\in \Gamma \left( F,\nu ,N\right) ,$ \emph{and}
\begin{equation*}
\begin{array}{l}
d^{F}\omega \left( z_{0},z_{1},...,z_{q}\right) =\overset{q}{\underset{i=0}{%
\tsum }}\left( -1\right) ^{i}\Gamma \left( \rho ,Id_{N}\right) z_{i}\left(
\omega \left( \left( z_{0},z_{1},...,\hat{z}_{i},...,z_{q}\right) \right)
\right) \\
~\ \ \ \ \ \ \ \ \ \ \ \ \ \ \ \ \ \ \ \ \ \ \ \ \ \ +\underset{i<j}{\tsum }%
\left( -1\right) ^{i+j}\omega \left( \left( \left[ z_{i},z_{j}\right]
_{F},z_{0},z_{1},...,\hat{z}_{i},...,\hat{z}_{j},...,z_{q}\right) \right) ,%
\end{array}%
\end{equation*}%
\emph{for any }$z_{0},z_{1},...,z_{q}\in \Gamma \left( F,\nu ,N\right) ,$
\emph{is unique having the following property:}%
\begin{equation*}
\begin{array}{c}
L_{z}=d^{F}\circ i_{z}+i_{z}\circ d^{F},~\forall z\in \Gamma \left( F,\nu
,N\right) .%
\end{array}%
\leqno(4.4)
\end{equation*}

This application is called\emph{\ the exterior differentiation ope\-ra\-tor
of the exterior differential algebra of the Lie algebroid }$((F,\nu
,N),[,]_{F},(\rho ,Id_{N})).$

\textbf{Theorem 4.5} \emph{The exterior differentiation operator }$d^{F}$%
\emph{\ given by the previous theorem has the following properties:}\medskip

\noindent 1. \emph{\ For any }$\omega \in $\emph{\ }$\Lambda ^{q}\left(
F,\nu ,N\right) $\emph{\ and }$\theta \in $\emph{\ }$\Lambda ^{r}\left(
F,\nu ,N\right) $\emph{\ we obtain }%
\begin{equation*}
\begin{array}{c}
d^{F}\left( \omega \wedge \theta \right) =d^{F}\omega \wedge \theta +\left(
-1\right) ^{q}\omega \wedge d^{F}\theta .%
\end{array}%
\leqno(4.5)
\end{equation*}

\noindent 2.\emph{\ For any }$z\in \Gamma \left( F,\nu ,N\right) $ we obtain
\begin{equation*}
\begin{array}{c}
L_{z}\circ d^{F}=d^{F}\circ L_{z}.%
\end{array}%
\leqno(4.6)
\end{equation*}

\noindent3.\emph{\ }$d^{F}\circ d^{F}=0.$

\bigskip \textbf{Theorem 4.6} (of Maurer-Cartan type)\textbf{\ \ }

\emph{If }$((F,\nu ,N),[,]_{F},(\rho ,Id_{N}))$ \emph{is a Lie algebroid and}
$d^{F}$\break \emph{is the ex\-te\-rior differentiation operator of the
exterior differential} $\mathcal{F}(N)$\emph{-algebra}\break $(\Lambda
(F,\nu ,N),+,\cdot ,\wedge ),$ \emph{then we obtain the structure equations
of Maurer-Cartan type }%
\begin{equation*}
\begin{array}{c}
d^{F}t^{\alpha }=-\displaystyle\frac{1}{2}L_{\beta \gamma }^{\alpha
}t^{\beta }\wedge t^{\gamma },~\alpha \in \overline{1,p}%
\end{array}%
\leqno(\mathcal{C}_{1})
\end{equation*}%
\emph{and\ }%
\begin{equation*}
\begin{array}{c}
d^{F}x^{i}=\rho _{\alpha }^{i}t^{\alpha },~i\in \overline{1,n},%
\end{array}%
\leqno(\mathcal{C}_{2})
\end{equation*}%
\emph{where }$\left\{ t^{\alpha },\alpha \in \overline{1,p}\right\} ~$\emph{%
is the coframe of the vector bundle }$\left( F,\nu ,N\right) .$\bigskip
\noindent

This equations will be called \emph{the structure equations of Maurer-Cartan
type associa\-ted to the Lie algebroid }$\left( \left( F,\nu ,N\right) ,%
\left[ ,\right] _{F},\left( \rho ,Id_{N}\right) \right) .$

\bigskip\noindent\emph{Proof.} Let $\alpha \in \overline{1,p}$ be arbitrary.
Since
\begin{equation*}
\begin{array}{c}
d^{F}t^{\alpha }\left( t_{\beta },t_{\gamma }\right) =-L_{\beta \gamma
}^{\alpha },~\forall \beta ,\gamma \in \overline{1,p}%
\end{array}%
\end{equation*}%
it results that
\begin{equation*}
\begin{array}{c}
d^{F}t^{\alpha }=-\underset{\beta <\gamma }{\tsum }L_{\beta \gamma }^{\alpha
}t^{\beta }\wedge t^{\gamma }.%
\end{array}%
\leqno(1)
\end{equation*}
Since $L_{\beta \gamma }^{\alpha }=-L_{\gamma \beta }^{\alpha }$ and $%
t^{\beta }\wedge t^{\gamma }=-t^{\gamma }\wedge t^{\beta }$, for nay $\beta
,\gamma \in \overline{1,p},$ it results that
\begin{equation*}
\begin{array}{c}
\underset{\beta <\gamma }{\tsum }L_{\beta \gamma }^{\alpha }t^{\beta }\wedge
t^{\gamma }=\displaystyle\frac{1}{2}L_{\beta \gamma }^{\alpha }t^{\beta
}\wedge t^{\gamma }%
\end{array}%
\leqno(2)
\end{equation*}%
Using the equalities $\left( 1\right) $ and $\left( 2\right) $ it results
the structure equation $(\mathcal{C}_{1}).$

Let $i\in \overline{1,n}$ be arbitrary. Since
\begin{equation*}
\begin{array}{c}
d^{F}x^{i}\left( t_{\alpha }\right) =\rho _{\alpha }^{i},~\forall \alpha \in
\overline{1,p}%
\end{array}%
\end{equation*}%
it results the structure equation $(\mathcal{C}_{2}).$\hfill \emph{q.e.d.}%
\bigskip

\textbf{Remark 4.1 }In the particular case of the standard Lie algebroid
\begin{equation*}
((TN,\tau _{N},N),[,]_{TN},(Id_{TN},Id_{N}))
\end{equation*}
we obtain
\begin{equation*}
\begin{array}{c}
d^{TN}x^{i}=dx^{i},~i\in \overline{1,n},%
\end{array}%
\leqno(\mathcal{C}_{2}^{\prime })
\end{equation*}%
where $\left\{ dx^{i},~i\in \overline{1,n}\right\} ~$is the coframe of the
vector bundle $(TN,\tau _{N},N).$

As $d^{TN}\circ d^{TN}=0$ and $L_{jk}^{i}=0,$ for all $i,j,k\in \overline{1,n%
}$ \bigskip \noindent we obtain%
\begin{equation*}
\begin{array}{c}
d^{F}\left( dx^{i}\right) =0=-\displaystyle\frac{1}{2}L_{jk}^{i}dx^{j}\wedge
dx^{k},~i\in \overline{1,n}%
\end{array}%
\leqno(\mathcal{C}_{1}^{\prime })
\end{equation*}

This equations are the structure equations of Maurer-Cartan type
associa\-ted to the standard Lie algebroid $((TN,\tau
_{N},N),[,]_{TN},(Id_{TN},Id_{N})).$

\bigskip \noindent \textbf{Theorem 4.7 (}of Cartan type)\textbf{\ }\emph{Let
}$\left( E,\pi ,N\right) $\emph{\ be an IDS of the Lie algebroid }%
\begin{equation*}
\left( \left( F,\nu ,N\right) ,\left[ ,\right] _{F},\left( \rho
,Id_{N}\right) \right) .
\end{equation*}

\emph{If }$\left\{ \Theta ^{r+1},...,\Theta ^{p}\right\} $\emph{\ is a base
of the }$\mathcal{F}\left( N\right) $\emph{-submodule }$\left( \Gamma \left(
E^{0},\pi ^{0},N\right) ,+,\cdot \right) $\emph{, then the IDS }$\left(
E,\pi ,N\right) $\emph{\ is involutive if and only if it exists }%
\begin{equation*}
\Omega _{\beta }^{\alpha }\in \Lambda ^{1}\left( F,\nu ,N\right) ,~\alpha
,\beta \in \overline{r+1,p}
\end{equation*}%
\emph{so that}
\begin{equation*}
d^{F}\Theta ^{\alpha }=\Sigma _{\beta \in \overline{r+1,p}}\Omega _{\beta
}^{\alpha }\wedge \Theta ^{\beta }\in \mathcal{I}\left( \Gamma \left(
E^{0},\pi ^{0},N\right) \right) .
\end{equation*}

\emph{Proof. }Let $\left\{ S_{1},...,S_{r}\right\} $ be a base of the $%
\mathcal{F}\left( N\right) $-submodule $\left( \Gamma \left( E,\pi ,N\right)
,+,\cdot \right) $

Let $\left\{ S_{r+1},...,S_{p}\right\} \in \Gamma \left( F,\nu ,N\right) $
so that $\left\{ S_{1},...,S_{r},S_{r+1},...,S_{p}\right\} $ is a base of
the $\mathcal{F}\left( N\right) $-module
\begin{equation*}
\left( \Gamma \left( F,\nu ,N\right) ,+,\cdot \right) .
\end{equation*}

Let $\Theta ^{1},...,\Theta ^{r}\in \Gamma \left( \overset{\ast }{F},\overset%
{\ast }{\nu },N\right) $ so that $\left\{ \Theta ^{1},...,\Theta ^{r},\Theta
^{r+1},...,\Theta ^{p}\right\} $ is a base of the $\mathcal{F}\left(
N\right) $-module
\begin{equation*}
\left( \Gamma \left( \overset{\ast }{F},\overset{\ast }{\nu },N\right)
,+,\cdot \right) .
\end{equation*}

For any $a,b\in \overline{1,r}$ and $\alpha ,\beta \in \overline{r+1,p}$, we
have the equalities:%
\begin{equation*}
\begin{array}{ccc}
\Theta ^{a}\left( S_{b}\right) & = & \delta _{b}^{a} \\
\Theta ^{a}\left( S_{\beta }\right) & = & 0 \\
\Theta ^{\alpha }\left( S_{b}\right) & = & 0 \\
\Theta ^{\alpha }\left( S_{\beta }\right) & = & \delta _{\beta }^{\alpha }%
\end{array}%
\end{equation*}

We remark that the set of the $2$-forms%
\begin{equation*}
\left\{ \Theta ^{a}\wedge \Theta ^{b},\Theta ^{a}\wedge \Theta ^{\beta
},\Theta ^{\alpha }\wedge \Theta ^{\beta },~a,b\in \overline{1,r}\wedge
\alpha ,\beta \in \overline{r+1,p}\right\}
\end{equation*}%
is a base of the $\mathcal{F}\left( M\right) $-module $\left( \Lambda
^{2}\left( F,\nu ,N\right) ,+,\cdot \right) .$

Therefore, we have%
\begin{equation*}
d^{F}\Theta ^{\alpha }=\Sigma _{b<c}A_{bc}^{\alpha }\Theta ^{b}\wedge \Theta
^{c}+\Sigma _{b,\gamma }B_{b\gamma }^{\alpha }\Theta ^{b}\wedge \Theta
^{\gamma }+\Sigma _{\beta <\gamma }C_{\beta \gamma }^{\alpha }\Theta ^{\beta
}\wedge \Theta ^{\gamma },\leqno\left( 1\right)
\end{equation*}%
where, $A_{bc}^{\alpha },B_{b\gamma }^{\alpha }$ and $C_{\beta \gamma
}^{\alpha },~a,b,c\in \overline{1,r},~\alpha ,\beta ,\gamma \in \overline{%
r+1,p}$ are real local functions so that $A_{bc}^{\alpha }=-A_{cb}^{\alpha }$
and $C_{\beta \gamma }^{\alpha }=-C_{\gamma \beta }^{\alpha }.$

Using the formula%
\begin{equation*}
\begin{array}{cl}
d^{F}\Theta ^{\alpha }\left( S_{b},S_{c}\right) & =\Gamma \left( \rho
,Id_{N}\right) S_{b}\left( \Theta ^{\alpha }\left( S_{c}\right) \right)
-\Gamma \left( \rho ,Id_{N}\right) S_{c}\left( \Theta ^{\alpha }\left(
S_{b}\right) \right) \\
& -\Theta ^{\alpha }\left( \left[ S_{b},S_{c}\right] _{F}\right) ,%
\end{array}%
\leqno\left( 2\right)
\end{equation*}%
we obtain that
\begin{equation*}
A_{bc}^{\alpha }=-\Theta ^{\alpha }\left( \left[ S_{b},S_{c}\right]
_{F}\right) ,~\leqno\left( 3\right)
\end{equation*}%
for any $b,c\in \overline{1,r}$ and $\alpha \in \overline{r+1,p}.$

We admit that $\left( E,\pi ,N\right) $ is an involutive \emph{IDS} of the
Lie algebroid $\left( \left( F,\nu ,N\right) ,\left[ ,\right] _{F},\left(
\rho ,Id_{N}\right) \right) .$

As $\left[ S_{b},S_{c}\right] _{F}\in \Gamma \left( E,\pi ,N\right) ,$ for
any $b,c\in \overline{1,r},$ it results that $\Theta ^{\alpha }\left( \left[
S_{b},S_{c}\right] _{F}\right) =0,$ for any $b,c\in \overline{1,r}$ and $%
\alpha \in \overline{r+1,p}.$ Therefore, for any $b,c\in \overline{1,r}$ and
$\alpha \in \overline{r+1,p},$ we obtain $A_{bc}^{\alpha }=0$ and
\begin{equation*}
\begin{array}{ccl}
d^{F}\Theta ^{\alpha } & = & \Sigma _{b,\gamma }B_{b\gamma }^{\alpha }\Theta
^{b}\wedge \Theta ^{\gamma }+\frac{1}{2}C_{\beta \gamma }^{\alpha }\Theta
^{\beta }\wedge \Theta ^{\gamma } \\
& = & \left( B_{b\gamma }^{\alpha }\Theta ^{b}+\frac{1}{2}C_{\beta \gamma
}^{\alpha }\Theta ^{\beta }\right) \wedge \Theta ^{\gamma }.%
\end{array}%
\end{equation*}

As
\begin{equation*}
\Omega _{\gamma }^{\alpha }\overset{put}{=}B_{b\gamma }^{\alpha }\Theta ^{b}+%
\frac{1}{2}C_{\beta \gamma }^{\alpha }\Theta ^{\beta }\in \Lambda ^{1}\left(
F,\nu ,N\right) ,~
\end{equation*}%
for any $\alpha ,\beta \in \overline{r+1,p},$ it results the first
implication.

Conversely, we admit that it exists
\begin{equation*}
\Omega _{\beta }^{\alpha }\in \Lambda ^{1}\left( F,\nu ,N\right) ,~\alpha
,\beta \in \overline{r+1,p}
\end{equation*}%
so that
\begin{equation*}
d^{F}\Theta ^{\alpha }=\Sigma _{\beta \in \overline{r+1,p}}\Omega _{\beta
}^{\alpha }\wedge \Theta ^{\beta },~\leqno\left( 4\right)
\end{equation*}%
for any $\alpha \in \overline{r+1,p}.$

Using the affirmations $\left( 1\right) ,\left( 2\right) $ and $\left(
4\right) $ we obtain that $A_{bc}^{\alpha }=0,~$for any $b,c\in \overline{1,r%
}$ and $\alpha \in \overline{r+1,p}.$

Using the affirmation $\left( 3\right) $, we obtain $\Theta ^{\alpha }\left( %
\left[ S_{b},S_{c}\right] _{F}\right) =0,~$for any $b,c\in \overline{1,r}$
and $\alpha \in \overline{r+1,p}.$

Therefore, we have $\left[ S_{b},S_{c}\right] _{F}\in \Gamma \left( E,\pi
,N\right) ,~$for any $b,c\in \overline{1,r}.$ Using the \emph{Proposition
3.2.2}, we obtain the second implication.\hfill \emph{q.e.d.}\medskip

\section{Exterior Differential Systems}

Let $\left( \left( F,\nu ,N\right) ,\left[ ,\right] _{F},\left( \rho
,Id_{N}\right) \right) $ be a Lie algebroid.

\textbf{Definition 5.1 }Any ideal $\left( \mathcal{I},+,\cdot \right) $ of
the exterior differential algebra of the Lie algebroid $\left( \left( F,\nu
,N\right) ,\left[ ,\right] _{F},\left( \rho ,Id_{M}\right) \right) $ closed
under differentiation operator $d^{F}$ $,$ namely $d^{F}\mathcal{I\subseteq I%
},$ is called \emph{differential ideal of the Lie algebroid }$\left( \left(
F,\nu ,N\right) ,\left[ ,\right] _{F},\left( \rho ,Id_{M}\right) \right) .$

\textbf{Definition 5.2 }Let $\left( \mathcal{I},+,\cdot \right) $ be a
differential ideal of the Lie algebroid
\begin{equation*}
\left( \left( F,\nu ,N\right) ,\left[ ,\right] _{F},\left( \rho
,Id_{M}\right) \right) .
\end{equation*}

If it exists an \emph{IDS} $\left( E,\pi ,N\right) $ so that for all $k\in
\mathbb{N}^{\ast }$ and $\omega \in \mathcal{I}\cap \Lambda ^{k}\left( F,\nu
,N\right) $ we have $\omega \left( u_{1},...,u_{k}\right) =0,$ for any $%
u_{1},...,u_{k}\in \Gamma \left( E,\pi ,N\right) ,$ then we will say that $%
\left( \mathcal{I},+,\cdot \right) $\emph{\ is an exterior differential
system (EDS) of the Lie algebroid }$\left( \left( F,\nu ,N\right) ,\left[ ,%
\right] _{F},\left( \rho ,Id_{N}\right) \right) .$

\textbf{Theorem 5.1 (}of Cartan type) \emph{The IDS }$\left( E,\pi ,N\right)
$\emph{\ of the Lie algebroid }%
\begin{equation*}
\left( \left( F,\nu ,N\right) ,\left[ ,\right] _{F},\left( \rho
,Id_{N}\right) \right)
\end{equation*}%
\emph{\ is involutive, if and only if the ideal generated by the }$\mathcal{F%
}\left( N\right) $\emph{-submodule }$\left( \Gamma \left( E^{0},\pi
^{0},N\right) ,+,\cdot \right) $\emph{\ is an EDS of the Lie algebroid }$%
\left( \left( F,\nu ,N\right) ,\left[ ,\right] _{F},\left( \rho
,Id_{N}\right) \right) .$

\emph{Proof. }Let $\left( E,\pi ,N\right) $ be an involutive \emph{IDS} of
the Lie algebroid
\begin{equation*}
\left( \left( F,\nu ,N\right) ,\left[ ,\right] _{F},\left( \rho
,Id_{N}\right) \right) .
\end{equation*}

Let $\left\{ \Theta ^{r+1},...,\Theta ^{p}\right\} $ be a base of the $%
\mathcal{F}\left( N\right) $-submodule $\left( \Gamma \left( E^{0},\pi
^{0},N\right) ,+,\cdot \right) .$

We know that
\begin{equation*}
\mathcal{I}\left( \Gamma \left( E^{0},\pi ^{0},N\right) \right) =\cup _{q\in
\mathbb{N}}\left\{ \Omega _{\alpha }\wedge \Theta ^{\alpha },~\left\{ \Omega
_{r+1},...,\Omega _{p}\right\} \subset \Lambda ^{q}\left( F,\nu ,N\right)
\right\} .
\end{equation*}

Let $q\in \mathbb{N}$ and $\left\{ \Omega _{r+1},...,\Omega _{p}\right\}
\subset \Lambda ^{q}\left( F,\nu ,N\right) $ be arbitrary.

Using the \emph{Theorems 4.5 and 4.7} we obtain
\begin{equation*}
\begin{array}{ccl}
d^{F}\left( \Omega _{\alpha }\wedge \Theta ^{\alpha }\right) & = &
d^{F}\Omega _{\alpha }\wedge \Theta ^{\alpha }+\left( -1\right) ^{q+1}\Omega
_{\beta }\wedge d^{F}\Theta ^{\beta } \\
& = & \left( d^{F}\Omega _{\alpha }+\left( -1\right) ^{q+1}\Omega _{\beta
}\wedge \Omega _{\alpha }^{\beta }\right) \wedge \Theta ^{\alpha }.%
\end{array}%
\end{equation*}

As
\begin{equation*}
d^{F}\Omega _{\alpha }+\left( -1\right) ^{q+1}\Omega _{\beta }\wedge \Omega
_{\alpha }^{\beta }\in \Lambda ^{q+2}\left( F,\nu ,N\right)
\end{equation*}%
it results that
\begin{equation*}
d^{F}\left( \Omega _{\beta }\wedge \Theta ^{\beta }\right) \in \mathcal{I}%
\left( \Gamma \left( E^{0},\pi ^{0},N\right) \right)
\end{equation*}

Therefore,
\begin{equation*}
d^{F}\mathcal{I}\left( \Gamma \left( E^{0},\pi ^{0},N\right) \right)
\subseteq \mathcal{I}\left( \Gamma \left( E^{0},\pi ^{0},N\right) \right) .
\end{equation*}

Conversely, let $\left( E,\pi ,N\right) $ be an \emph{IDS} of the Lie
algebroid $\left( \left( F,\nu ,N\right) ,\left[ ,\right] _{F},\left( \rho
,Id_{N}\right) \right) $ so that the $\mathcal{F}\left( N\right) $-submodule
$\left( \mathcal{I}\left( \Gamma \left( E^{0},\pi ^{0},N\right) \right)
,+,\cdot \right) $ is an \emph{EDS} of the Lie algebroid $\left( \left(
F,\nu ,N\right) ,\left[ ,\right] _{F},\left( \rho ,Id_{N}\right) \right) .$

Let $\left\{ \Theta ^{r+1},...,\Theta ^{p}\right\} $ be a base of the $%
\mathcal{F}\left( N\right) $-submodule $\left( \Gamma \left( E^{0},\pi
^{0},N\right) ,+,\cdot \right) .$ As
\begin{equation*}
d^{F}\mathcal{I}\left( \Gamma \left( E^{0},\pi ^{0},N\right) \right)
\subseteq \mathcal{I}\left( \Gamma \left( E^{0},\pi ^{0},N\right) \right)
\end{equation*}%
it results that it exists
\begin{equation*}
\Omega _{\beta }^{\alpha }\in \Lambda ^{1}\left( F,\nu ,N\right) ,~\alpha
,\beta \in \overline{r+1,p}
\end{equation*}%
so that
\begin{equation*}
d^{F}\Theta ^{\alpha }=\Sigma _{\beta \in \overline{r+1,p}}\Omega _{\beta
}^{\alpha }\wedge \Theta ^{\beta }\in \mathcal{I}\left( \Gamma \left(
E^{0},\pi ^{0},N\right) \right) .
\end{equation*}

Using the \emph{Theorem 4.7} there results that $\left( E,\pi ,N\right) $ is
an involutive \emph{IDS.}\hfill \emph{q.e.d.}\medskip\ \

\bigskip

\section*{Acknowledgment}

\addcontentsline{toc}{section}{Acknowledgment}

I would like to thank Matsumae International Foundation for the research
grant at Tokai University, during April-September 2008. I would also like to
thank Professors Hideo SHIMADA and Sorin Vasile SABAU from Tokai
University-Japan, for useful discussions and their suggestions.

\bigskip \ \ \ \ 




\end{document}